\documentclass[prl,preprint,endfloats]{revtex4}

\usepackage{graphicx}
\usepackage{bm}
\usepackage{pifont}
\usepackage{amsmath}

\begin{document}

\title{Direct visualization of three-dimensional shape of skyrmion strings in a noncentrosymmetric magnet}

\author{S. Seki$^{1,2,3,4,*}$, M. Suzuki$^{5,*}$, M. Ishibashi$^6$, R. Takagi$^{1,2,3}$, N. D. Khanh$^3$, Y. Shiota$^6$, W. Koshibae$^{3}$, Y. Tokura$^{1,3,7}$, T. Ono$^{6,8,*}$} 
\affiliation{$^1$ Department of Applied Physics, University of Tokyo, Tokyo 113-8656, Japan, \\ $^2$ Institute of Engineering Innovation, University of Tokyo, Tokyo 113-8656, Japan, \\ $^3$ RIKEN Center for Emergent Matter Science (CEMS), Wako 351-0198, Japan, \\ $^4$ PRESTO, Japan Science and Technology Agency (JST), Kawaguchi 332-0012, Japan, \\ $^5$ Japan Synchrotron Radiation Research Institute, Sayo 679-5198, Japan, \\ $^6$ Institute for Chemical Research, Kyoto University, Uji 611-0011, Japan, \\ $^7$ Tokyo College, University of Tokyo, Tokyo 113-8656, Japan, \\ $^8$ Center for Spintronics Research Network, Graduate School of Engineering Science, Osaka University, Toyonaka 560-853, Japan}

\begin{abstract}

{\bf Magnetic skyrmion, i.e. a topologically stable swirling spin texture, appears as a particle-like object in the two-dimensional (2D) systems, and has recently attracted attention as a candidate of novel information carrier. In the real three-dimensional (3D) systems, a skyrmion is expected to form a string structure along an extra dimension, while its experimental identification has rarely been achieved. Here, we report the direct visualization of 3D shape of individual skyrmion strings, for the recently discovered room-temperature skyrmion-hosting noncentrosymmetric compound Mn$_{1.4}$Pt$_{0.9}$Pd$_{0.1}$Sn. For this purpose, we have newly developed the magnetic X-ray tomography measurement system that can apply magnetic field, which plays a key role on the present achievement. Through the tomographic reconstruction of the 3D magnetization distribution based on the transmission images taken from various angles, a genuine skyrmion string running through the entire thickness of the sample, as well as various defect structures such as the interrupted and Y-shaped strings, are successfully identified. The observed point defect may represent the emergent magnetic monopole, as recently proposed theoretically. The present tomographic approach with tunable magnetic field paves the way for the direct visualization of the structural dynamics of individual skyrmion strings in the 3D space, which will contribute to the better understanding of the creation, annihilation and transfer process of these topological objects toward the potential device applications.}

\end{abstract}

\maketitle

Magnetic skyrmion appears as a vortex-like swirling spin texture in the 2D systems, which has the character of topologically stable particle\cite{MnSi, TEMFeCoSi, SkXTheoryFirst, SkXReviewFertTwo, SkXReviewTokura}. In metallic materials, skyrmions can be moved by five order of magnitude smaller electric current than the conventional ferromagnetic domain wall\cite{CurrentControlNeutron, CurrentControlLTEM, CurrentControlBubble, CurrentControlSTXM}, and the associated emergent electromagnetic fields originating from quantum Berry phase also causes various nontrivial transport phenomena\cite{SkXReviewTokura, THE, EmergentEfield}. Because of such particle-like nature and unique electric controllability, skyrmions are now intensively studied as the potential information carrier for magnetic information storage and processing devices\cite{SkXReviewFertTwo, SkLogic, Neuromorphic}.

In the 3D systems, the skyrmion is predicted to form a string structure (Figs. 1a and b), which consists of the uniform stacking of 2D skyrmion spin texture\cite{Monopole}. Such a string-like feature is particularly important when we discuss the creation, annihilation and motion of skyrmions induced by external stimuli such as electric current, since these processes should be deeply related to the shape deformation and pinning/depinning of string structures\cite{CurrentControl3D, CurrentControl3D2, CurrentControl3D3, SkStringMagnon}. Therefore, toward the efficient manipulation of skyrmions, the establishment of new experimental approaches to visualize the 3D shape of individual skyrmion string is essential.

In general, magnetic skyrmions are stabilized in noncentrosymmetric systems, where the Dzyaloshinskii-Moriya (DM) interaction and external magnetic field $B$ plays an important role on the skyrmion formation\cite{SkXReviewFertTwo, SkXReviewTokura}. As shown in Fig. 1a, the local magnetic moment at the core and edge of each skyrmion is antiparallel and parallel to $B$, respectively. At the intermediate region, the in-plane component of magnetic moment shows various swirling textures depending on the symmetry of underlying crystallographic lattice, and Bloch\cite{MnSi, TEMFeCoSi, Cu2OSeO3_Seki, CoZnMn_First}, N\'{e}el\cite{SkXReviewFertTwo, GaV4S8} and anti-vortex\cite{Heusler} type skyrmions appear in the system with chiral, polar and $D_{2d}$ symmetry, respectively. The orientation of skyrmion string is usually aligned along the external magnetic field direction.

In such noncentrosymmetric compounds, the skyrmion formation has previously been identified through the neutron and X-ray scattering in the reciprocal space\cite{MnSi, CurrentControlNeutron}, or the 2D imaging techniques in the real space, such as Lorentz transmission electron microscopy (LTEM)\cite{TEMFeCoSi, CurrentControlLTEM}, electron holography\cite{Holography}, magnetic force microscopy (MFM)\cite{Monopole, GaV4S8}, and scanning transmission X-ray microscopy (STXM)\cite{CurrentControlSTXM, FertSTXM}. In the latter real-space approaches, a thin plate-shaped sample is viewed from the out-of-plane direction, and skyrmions are usually identified as the particle-like circular pattern under the out-of-plane magnetic field. Very recently, STXM and LTEM measurement has also been performed under the in-plane magnetic field, where the stripe-like pattern corresponding to the skyrmion tube fragments embedded in a helical magnetic background has been observed\cite{MaxSTXM, YuSkTube}. In these previous experiments, however, the sample was always viewed from one fixed direction, and the information along the depth direction is averaged out and lost. Therefore, the detailed 3D shape of skyrmion string cannot be identified. Here, one potential solution is the employment of the tomographic approach, i.e. the procedure used for the CT (computed tomography) scan of human body. In this method, 2D transmission images are taken from various angles, and then the 3D structure of the target object can be reconstructed. It has recently been demonstrated that such a 3D tomographic imaging technique is applicable even for the spin systems\cite{TomographyNature, TomographySuzuki, TomographyFilm, TomographyReview}, while the previous measurements were carried out only for the simple centrosymmetric ferromagnets without magnetic field. Thus, the 3D imaging of skyrmion string, which is stabilized in special noncentrosymmetric materials under external magnetic field, remains an important challenge.

In this study, we have successfully visualized the 3D shape of individual skyrmion strings, by employing the recently discovered room-temperature skyrmion-hosting noncentrosymmetric material Mn$_{1.4}$Pt$_{0.9}$Pd$_{0.1}$Sn\cite{Heusler}. For this purpose, we have newly developed the magnetic tomography measurement system that can apply magnetic field, which plays a key role for the present achievements. In addition to the genuine skyrmion strings running through the entire thickness of the sample, various defect structures such as the interrupted and Y-shaped strings have been identified, the latter of which can be accompanied with the monopole of emergent magnetic flux\cite{Monopole, MonopoleTheory}. Our results suggest that the magnetic 3D tomographic imaging with tunable magnetic field can be a powerful tool for the better understanding of the formation process and dynamics of skyrmion strings.

Our target compound Mn$_{1.4}$Pt$_{0.9}$Pd$_{0.1}$Sn is known as a rare example of room-temperature skyrmion-hosting compounds\cite{Heusler}, and is characterized by the inverse Heusler crystal structure with noncentrosymmetric tetragonal $D_{2d}$ symmetry (space group $I\bar{4}2d$). Magnetic ordering temperature is around 400 K, and the easy-axis of magnetization is along the [001] direction. By applying $B \parallel [001]$, the anti-vortex type skyrmions are stabilized at room temperature, where the skyrmion diameter depends on the sample thickness according to the previous reports\cite{Heusler, HeuslerAnisotropy, HeuslerMFM}. 

Experimental setup for the magnetic X-ray tomography and the associated scanning X-ray magnetic circular dichroism (XMCD) measurement is shown in Figs. 1c and d\cite{TomographySuzuki}. Here, the direction of incident X-ray beam is defined as the $Y$-axis. X-ray energy was tuned at 11.572 keV, in resonance with the $L_3$ absorption edge of Pt (Fig. S1). The wedge-shaped single crystal of Mn$_{1.4}$Pt$_{0.9}$Pd$_{0.1}$Sn (Figs. 1e and f) was placed at the focal point of X-ray beam with the spot size of $150 \times 150$ nm$^2$ in full width at half maximum (FWHM), and the absorption coefficient ($\mu=-\ln(I_{\rm t}/I_{\rm 0})$ with $I_{\rm t}$ and $I_{\rm 0}$ being the intensity of the transmitted and incident X-ray beam, respectively) for the left- and right-handed circular polarized beam ($\mu^+$ and $\mu^-$, respectively) was measured by the silicon photodiode detector. The amplitude of XMCD $\Delta \mu = \mu^+ - \mu^-$ reflects the $Y$-component of local magnetization integrated over the beam path, and the polarization-averaged absorption $\bar{\mu} = (\mu^+ + \mu^-)/2$ is proportional to the electron density of the sample. The sample and the electromagnet are mounted on the common pulse-motor stages with a feedback control, which enables the translation and rotation of the sample keeping the magnetic field applied along the [001] axis of Mn$_{1.4}$Pt$_{0.9}$Pd$_{0.1}$Sn crystal (See Supplementary Note II for the details). By scanning the sample position within the $XZ$-plane, the 2D XMCD image is obtained. As shown in Fig. 1d, the orientation of the sample and the magnetic field can be simultaneously rotated around the $Z$-axis, and the angle between the incident X-ray direction (i.e. $Y$-axis) and the [001] axis of the sample is defined as $\theta$. Based on the 2D XMCD images taken at various angles $\theta$, the tomographic reconstruction of the magnetization distribution has been performed (See the Methods section and Ref. \cite{TomographySuzuki}). All the measurements were performed at BL39XU of the SPring-8 synchrotron radiation facility.

In Figs. 1e and f, the detailed shape of the present Mn$_{1.4}$Pt$_{0.9}$Pd$_{0.1}$Sn sample, prepared from a bulk single crystal by the FIB (focused ion beam) micro-fabrication technique, is indicated. The sample has the thin-plate shape with the thickness gradient, and the widest face is parallel to the (001) plane. Figure 2h indicates the SEM (scanning electron microscope) image of the sample viewed from the [001] orientation. The corresponding 2D X-ray absorption ($\bar{\mu}$) image, taken by scanning the position of incident X-ray beam normal to the sample plane (i.e. $\theta = 0$), is indicated in Fig. 2i. The line-scan profile in Fig. 2j confirms the smooth thickness gradient of the sample along the $Z$-direction. In this $\theta = 0$ configuration, we investigated the development of XMCD images (i.e., 2D distribution of magnetization projected along the X-ray beam direction) with increasing magnetic field applied along the out-of-plane [001] direction parallel to the X-ray beam. 

Figure 2a is the XMCD image with $B=0$, taken for the same region as Fig. 2i. Here, the XMCD contrast reflects the amplitude of out-of-plane magnetization component (i.e., the [001] component of local magnetization $m^c$ for this setup). The observed stripe pattern suggests the formation of helical magnetic order, where neighboring spins rotate within a plane normal to the magnetic modulation vector $q \parallel \langle100\rangle$\cite{Heusler}. As the sample thickness increases, the magnetic modulation period is found to be longer and reaches 1 $\mu$m at the thickest part of the sample. By applying the out-of-plane $B \parallel [001]$, the helical magnetic stripes gradually turn into the circular magnetic skyrmions (Figs. 2b-g), whose cores are characterized by the negative sign of $m^c$ antiparallel to $B$. These results are consistent with the recent reports based on MFM measurements\cite{HeuslerMFM, PtMnGa}. Figure 2k indicates the magnetic field dependence of XMCD intensity taken with the larger beam spot size $ \sim 13$ $\mu$m (corresponding to bulk magnetization averaged over the sample), which confirms that the saturated uniform ferromagnetic state is obtained at 500 mT. 

Next, we focused on the selected region of Fig. 2f (surrounded by dashed lines) in the skyrmion phase at 437 mT, and performed the same scanning XMCD measurements for different $\theta$ angles while keeping the $B \parallel [001]$ unchanged. In general, a cylindrical skyrmion string (Fig. 1a) gives a circular magnetic contrast when viewed from the string direction, but will provide more elongated one when viewed from the oblique angles. Indeed, the circular XMCD pattern observed at $\theta = 0$ (Fig. 3a) is gradually elongated along the $X$-direction as the tilting angle $\theta$ increases (Figs. 3b-e), which implies the validity of the cylindrical skyrmion string picture. We performed the same measurements for $-90^\circ \leq \theta \leq 90^\circ$ with $5^\circ$ steps, and the obtained XMCD images are used for the tomographic reconstruction of magnetization distribution. (A series of XMCD images taken from various angles are summarized as an animation in Supplementary Video 1.)

In general, the isolated magnetic skyrmion is characterized by the negative sign of core magnetization antiparallel to $B \parallel [001]$, which is embedded in the uniform ferromagnetic background with magnetization parallel to $B$ (Fig. 1a). To identify the 3D shape of magnetic skyrmion, we focus on the spatial distribution of the [001] component of local magnetization $m^c (\bf {r})$ and neglect the other magnetization component. Here, the observed XMCD image is taken in the $X$-$Y$-$Z$ coordinates fixed to the measurement system, while magnetization distribution is defined in the coordinates ${\bf r}=Z{\bf a} + x{\bf b} + y{\bf c}$ fixed to the crystallographic axes of the sample (Fig. 1d). For the specific value of $Z$ and $\theta$, the relation between the magnetization distribution $m^c (x,y)$ and the observed XMCD contrast $\Delta \mu (X, \theta)$ is given by
\begin{equation}
\frac{\Delta \mu (X, \theta)}{\cos \theta} = \int m^c(x, y) dY,
\label{RadonEq}
\end{equation}
which corresponds to the Radon transform of $m^c(x, y)$\cite{Radon}. Therefore, based on a series of XMCD images and standard tomographic reconstruction algorithm, the 3D spatial distribution of $m^c ({\bf r})$ can be straightforwardly obtained\cite{TomographySuzuki, TomographyMath}. In Fig. S3, the experimentally measured XMCD image and the one simulated from the reconstructed $m^c ({\bf r})$ and Eq. (\ref{RadonEq}) are indicated for the selected $\theta$ values. The measured and reproduced images are in good agreement with each other, which confirms the reliability of the present tomographic reconstruction process.

Now, we discuss the detail of the experimentally reconstructed $m^c ({\bf r})$ in the 3D space. Figure 4a indicates the XMCD image viewed from the [001] axis, where circular skyrmion cores with negative sign of $m^c$ are observed. Figure 4b represents a cross-section of the reconstructed $m^c ({\bf r})$ profile along the line I in Fig. 4a, cutting a well-defined skyrmion core. The negative $m^c$ region is found to form a straight line along the [001] direction connecting the top and bottom surfaces. Figure 4d indicates the 3D oblique view of reconstructed $m^c ({\bf r})$ profile, which clearly visualizes the rod-shaped strings aligned along the vertical direction. The above results demonstrate that a magnetic skyrmion in bulk crystals indeed appears in the form of a cylindrical string running through the entire thickness of the sample. (Supplementary Videos 2 and 3 provide the 360$^\circ$ view and slice animation of reconstructed $m^c ({\bf r})$ profiles, respectively, which reveals the 3D shape of individual skyrmion strings in more detail.)

According to the latest theories, skyrmion strings are also predicted to host several different types of defect structures\cite{Monopole, MonopoleTheory, BobberTheory}. In Fig. 4c, another cross-sectional image along the line II in Fig. 4a is presented. In the left side, the skyrmion string with negative $m^c$ forms a straight line starting from the top surface, while it is terminated at the middle of the sample. This interrupted skyrmion string is characterized by the weaker magnetic contrast than the genuine one in the top-view XMCD image (Fig. 4a), reflecting its limited string length. On the other hand, in the right side of Fig. 4c, two parallel skyrmion strings starting from the top surface are found to merge into a single string connected to the bottom surface. Such interrupted or Y-shaped string structures have recently been proposed in Refs. \cite{Monopole, MonopoleTheory}, the former of which is often referred to as the chiral bobber and considered as another candidate of information carrier\cite{BobberTheory, BobberLTEM}. The present results provide a direct experimental evidence to prove the existence of these predicted defect structures in the 3D space.

Theoretically, conduction electrons interacting with such spin textures should feel the emergent magnetic field ${\bf b}^{\rm em}({\bf r})$ associated with quantum Berry phase\cite{SkXReviewTokura}. In case of a genuine skyrmion string, the individual skyrmion string carries a quantized emergent magnetic flux along the direction of string. On the other hand, for the interrupted or Y-shaped skyrmion strings, the divergence of emergent magnetic flux $\nabla \cdot {\bf b}^{\rm em} \neq 0$ must appear at the termination or merging point of skyrmion strings. These singular point defects are referred to as the emergent magnetic monopoles (or anti-monopoles)\cite{Monopole, MonopoleTheory}. Here, we performed the micromagnetic simulation by assuming the $D_{2d}$ symmetry of DM interaction, and estimated the local magnetization distribution ${\bf m}({\bf r})$ for the genuine, interrupted and Y-shaped skyrmion strings as shown in Fig. 1b, Fig. 4e and Fig. 4f, respectively. The calculation of the corresponding emergent magnetic field distribution ${\bf b}^{\rm em}({\bf r})$ supports the existence of emergent magnetic monopole at termination or merging points of strings, whose positions are highlighted by the white circles in Figs. 4e and f (See Fig. S4 and Supplementary Note IV for the detail). Recently, the dynamics of such emergent magnetic monopoles has been considered as the key to interpret the nontrivial electron transport properties in topological magnetic phases\cite{MonopoleMR, MonopoleThermal}. The present approach provides a unique experimental method to track the 3D position of individual emergent magnetic monopoles under external magnetic field, which will contribute to the deeper analysis of the various exotic phenomena associated with emergent electromagnetic fields.

In this study, by developing the magnetic X-ray tomography measurement system with tunable magnetic field, we have successfully visualized the 3D shape of individual skyrmion strings and their defect structures at room temperature for a noncentrosymmetric Heusler compound. The present results pave the way for the direct 3D imaging of creation, annihilation as well as transfer process of individual skyrmion strings driven by the external stimuli, and will contribute to the better understanding of their intricate dynamics toward the potential device application. The above experiments also expand the reach of magnetic X-ray tomography in terms of both methodology (i.e. additional magnetic field environment) and target material system, and will promote the further advancement of this emerging 3D magnetic imaging technique. Because of the short wavelength of X-ray, the spatial resolution of the magnetic tomography can be further improved in principle. The full reconstruction of the vector magnetization distribution\cite{TomographyNature, VortexRing} (which is possible by two-axis rotation of the target object but currently successful only under zero magnetic field) and the direct experimental confirmation of winding number will be another important future challenge. Previously, the stability and dynamics of skyrmions have been studied mainly based on the 2D imaging and associated theoretical simulations. The present 3D visualization approach allows us to access to the unexplored third dimension of skyrmions, which may signal a new phase in the development of skyrmionics.

\begin{figure}
\begin{center}
\includegraphics*[width=11.7cm]{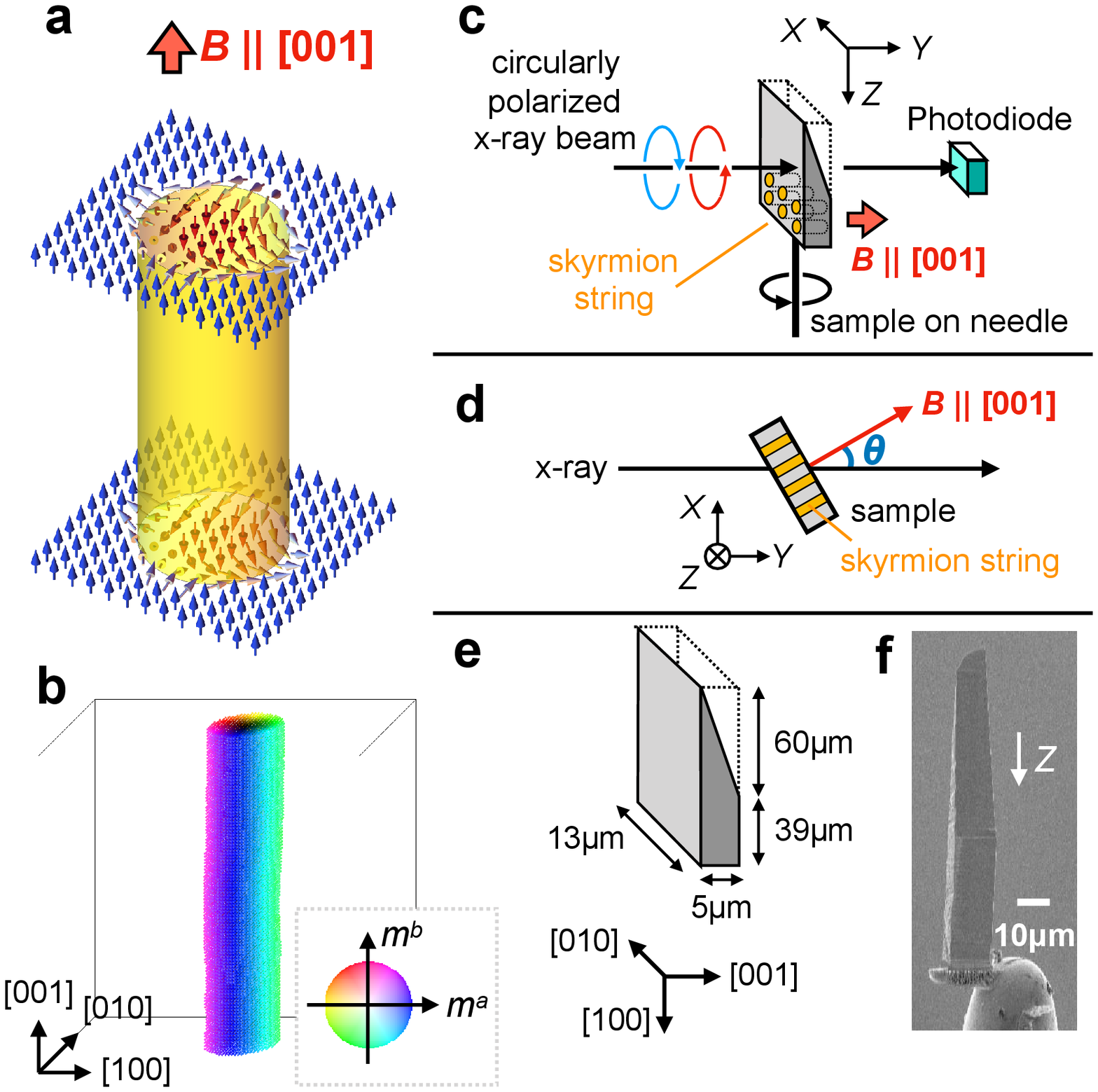}
\caption{{\bf Experimental setup for the magnetic X-ray tomography measurements.} {\bf a}, Schematic illustration of skyrmion string. {\bf b}, The corresponding spatial distribution of local magnetic moment ${\bf m}({\bf r})$, obtained by the micromagnetic simulation (See Supplementary Note IV for the detail). The hue and brightness of background color represent the in-plane ($m^a$ and $m^b$) and out-of-plane ($m^c$) component of ${\bf m}({\bf r})$, respectively. The length of magnetic moment is fixed to $|{\bf m} ({\bf r})|=1$. Note that we do not put color for $m^c > 0.5$., i.e. the outside of the skyrmion string, to make the three-dimensional magnetic texture clear. {\bf c}, Experimental setup for the magnetic X-ray tomography measurements. {\bf d}, The top view of {\bf c}. $\theta$ is defined as the angle between the X-ray beam direction and the [001] axis of Mn$_{1.4}$Pt$_{0.9}$Pd$_{0.1}$Sn single crystal. The sample and electromagnet are mounted on the same rotational stage, and magnetic field $B$ is always applied parallel to the [001] direction (See Supplementary Note II). {\bf e}, Schematic illustration showing the shape of the Mn$_{1.4}$Pt$_{0.9}$Pd$_{0.1}$Sn sample. {\bf f}, SEM image of the sample attached to the needle.}
\end{center}
\end{figure}

\begin{figure}
\begin{center}
\includegraphics*[width=17cm]{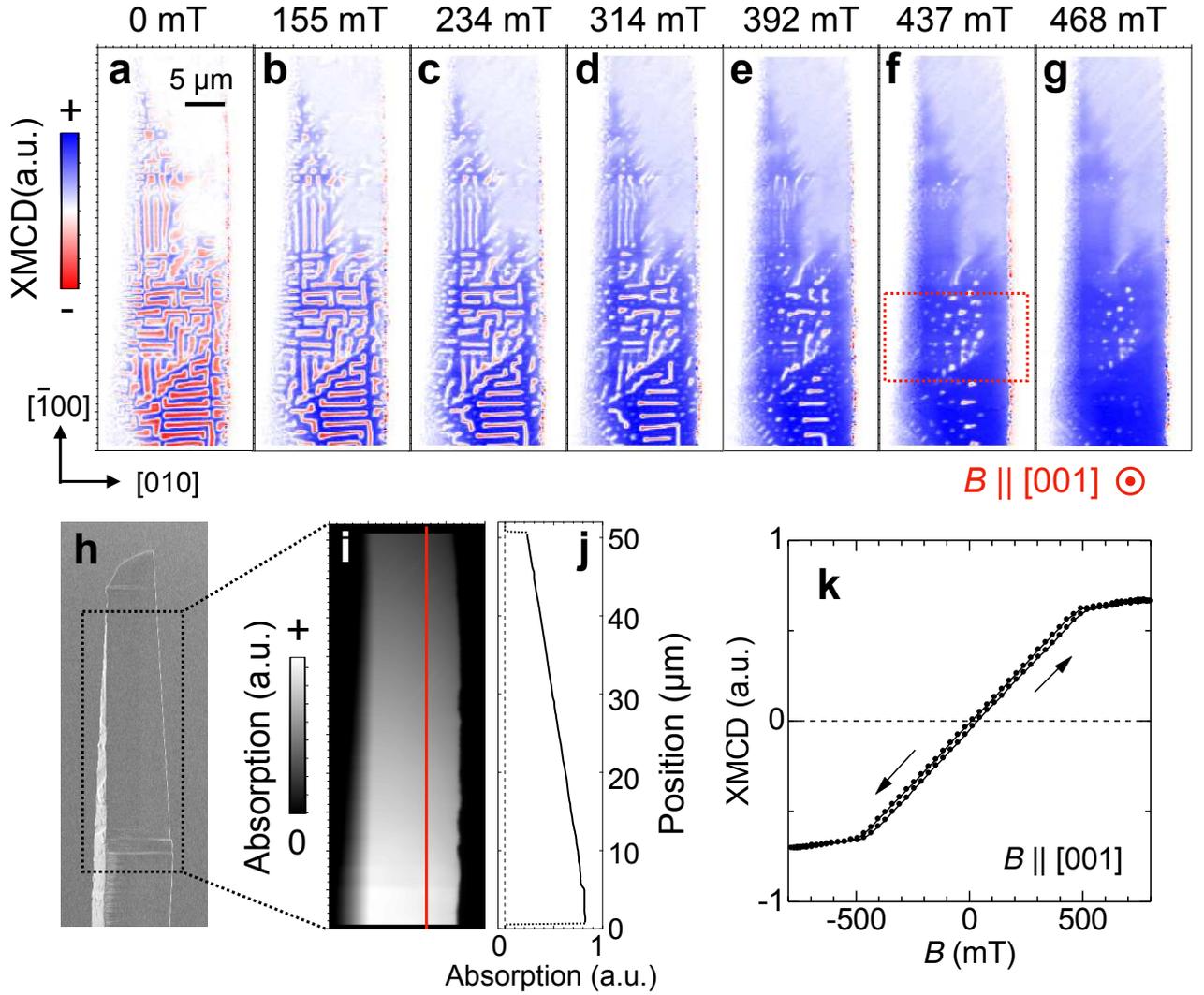}
\caption{{\bf Magnetic field dependence of XMCD pattern taken from the [001] direction.} {\bf a}-{\bf g}, XMCD images measured for various amplitude of out-of-plane magnetic field $B \parallel [001]$ with the $\theta = 0$ condition. The color represents the XMCD intensity $\Delta \mu$ reflecting the out-of-plane component of local magnetization. {\bf h}, SEM image of Mn$_{1.4}$Pt$_{0.9}$Pd$_{0.1}$Sn sample viewed from the [001] direction. {\bf i}, The corresponding X-ray absorption image taken for the boxed region in {\bf h}. XMCD images in {\bf a}-{\bf g} are also taken for the same region. The color represents the polarization-averaged absorption intensity $\bar{\mu}$, which is proportional to the sample thickness. {\bf j}, Position dependence of absorption intensity taken along the red line in {\bf i}. {\bf k}, Magnetic field dependence of XMCD intensity $\Delta \mu$ measured with the larger X-ray beam spot size, which corresponds to the averaged bulk magnetization.}
\end{center}
\end{figure}

\begin{figure}
\begin{center}
\includegraphics*[width=17cm]{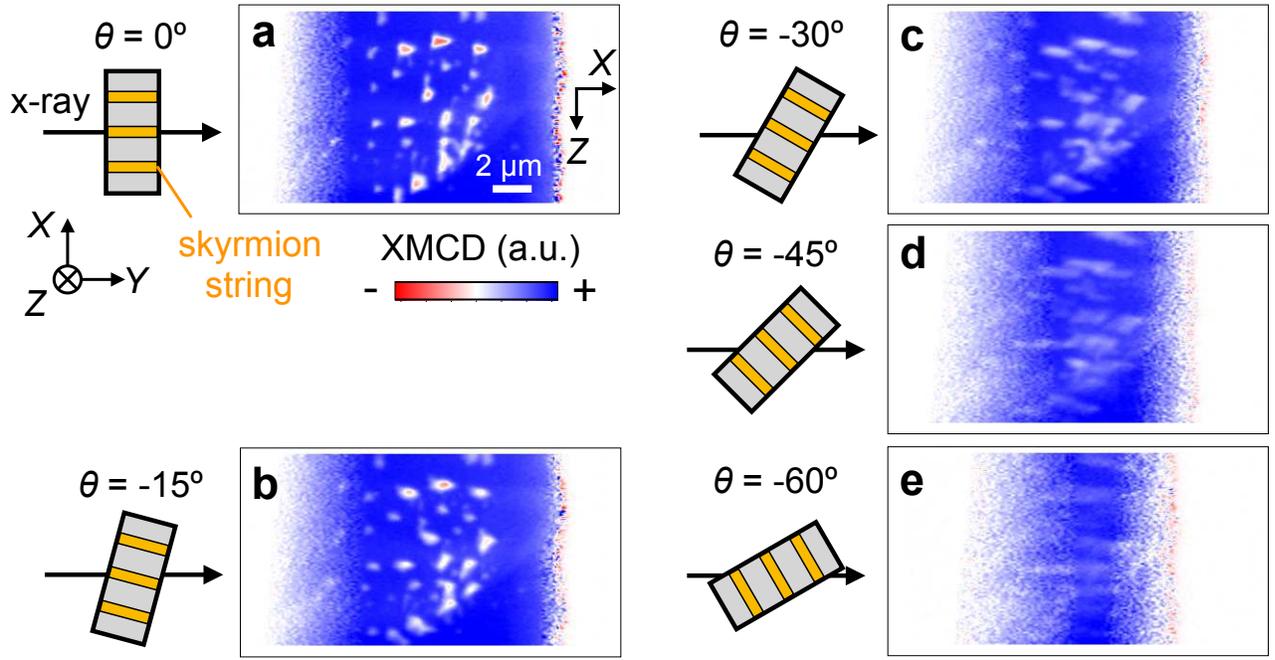}
\caption{{\bf XMCD images in the skyrmion state taken from various angles $\theta$.} {\bf a}-{\bf e}, XMCD images taken at 437 mT with various $\theta$ values. The measurement is performed for the boxed region in Fig. 2f, and the magnetic field direction is always fixed along the [001] direction of the Mn$_{1.4}$Pt$_{0.9}$Pd$_{0.1}$Sn sample (Fig. 1d). The color represents the XMCD intensity $\Delta \mu$ reflecting the $Y$-component of local magnetization integrated over the beam path. Schematic illustration of experimental configuration is shown at the left side of each data.}
\end{center}
\end{figure}

\begin{figure}
\begin{center}
\includegraphics*[width=17cm]{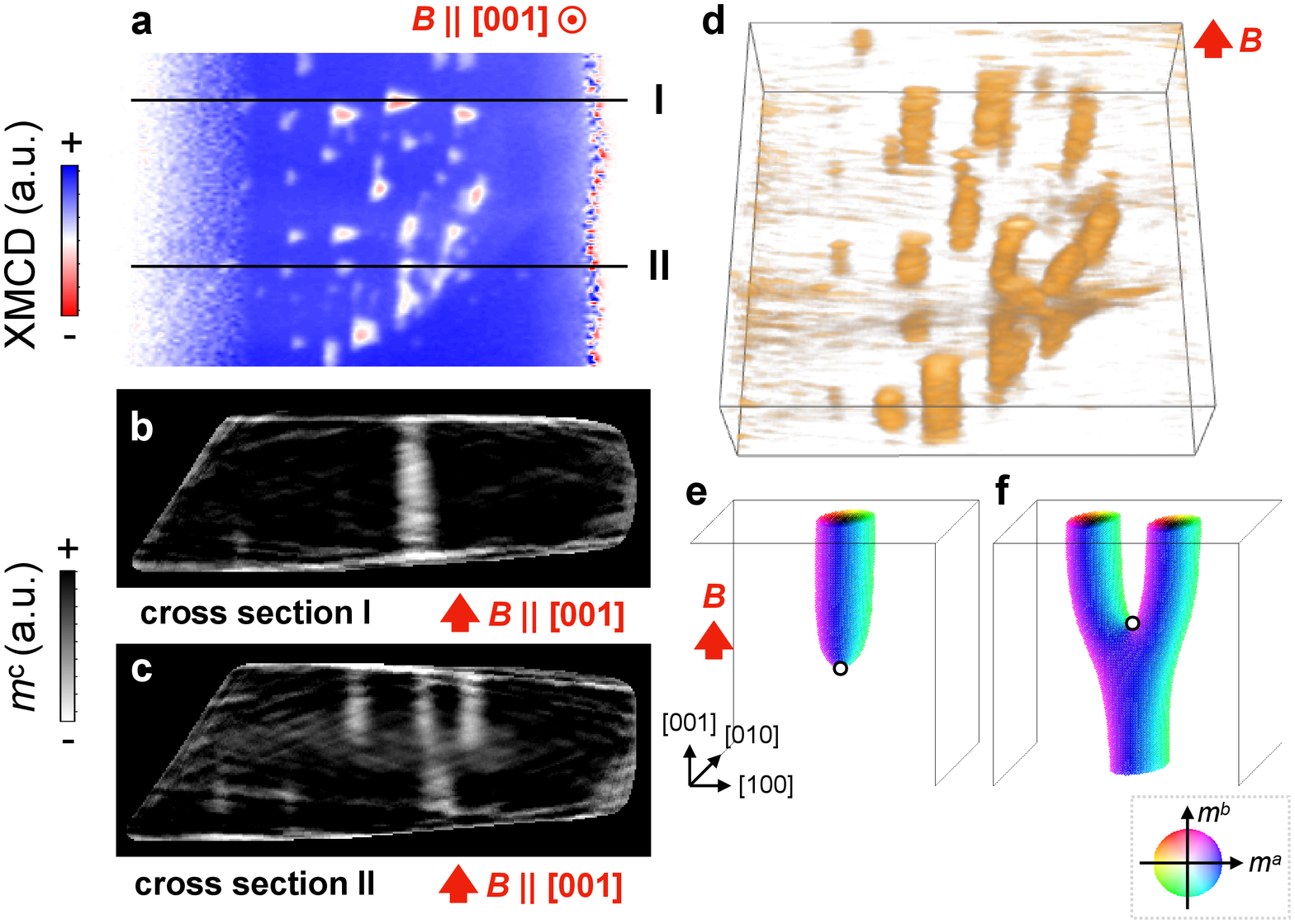}
\caption{{\bf 3D distribution of $m^c$ in the skyrmion state obtained by the tomographic reconstruction.} {\bf a}, XMCD image at 437 mT taken from the [001] direction, corresponding to the boxed region in Fig. 2f. {\bf b,c}, Cross section of experimentally reconstructed $m^c$({\bf r}) profile along the line I and II in {\bf a}, respectively. The color represents the [001] component of local magnetization $m^c$. {\bf d}, Oblique view of the experimentally reconstructed $m^c ({\bf r})$ profile, where the orange color represents the amplitude of negative $m^c$ component. {\bf e,f}, Spatial distribution of ${\bf m}({\bf r})$ for the interrupted and Y-shaped skyrmion string, obtained by the micromagnetic simulation. The color definition is the same as the one for Fig. 1b. The white circle represents the position of emergent magnetic monopole (See Supplementary Note IV and Fig. S4 for the detail).}
\end{center}
\end{figure}

\section*{Methods} 

\subsection*{Sample preparation and characterization.}

Polycrystalline sample of Mn$_{1.4}$Pt$_{0.9}$Pd$_{0.1}$Sn was prepared by the arc-melting stoichiometric amount of pure Mn, Pt, Pd, and Sn pieces under an Ar atmosphere. Bulk single crystals were grown by the slow-cooling of melted polycrystalline samples sealed into a silica tube under vacuum. Crystal orientations were determined using the back-reflection X-ray Laue photography method, and the purity of the sample was confirmed by the powder X-ray diffraction. The wedge-shaped single crystal sample for the X-ray tomography measurements was prepared by the focused-ion-beam (FIB) micro-fabrication technique, which was attached to a needle on the sample holder by tungsten bonding.

\subsection*{Magnetic X-ray tomography measurements.}

The scanning hard-X-ray microtomography experiment was conducted at BL39XU of the SPring-8\cite{TomographySuzuki, Method1}. X-ray radiation from the standard in-vacuum undulator was monochromatized using a Si (111) double-crystal monochromator. The X-ray energy was tuned at 11.572 keV, at which the present sample of Mn$_{1.4}$Pt$_{0.9}$Pd$_{0.1}$Sn showed the maximum amplitude of the XMCD spectrum at the Pt $L_3$ edge (See Fig. S1 and Supplementary Note I). A 1.4-mm-thick diamond X-ray phase plate was used to generate a circularly polarized X-ray beam with the degree of circular polarization of $P_C > 0.99$. The circularly-polarized monochromatic X-ray beam was then focused onto a sample with a spot size of $150 \times 150$ nm$^2$ in full width at half maximum (FWHM) using the elliptical mirrors in the Kirkpatrick-Baez configuration. The depth of the focus was 200 $\mu$m, which was much greater than the sample diameter and the eccentric radius of the sample rotation stage. 

Schematic illustration of the experimental setup is shown in Figs. 1c and d. The wedge-shaped Mn$_{1.4}$Pt$_{0.9}$Pd$_{0.1}$Sn sample was placed on the top of a projection-type electromagnet, and both were mounted on the same rotation stage ($\theta$) and two-dimensional translation ($X$-$Z$) stages. In this system, the sample and the electromagnet were rotated and moved while the relative orientation of the sample crystal axis and the magnetic field was unchanged during tomographic data acquisition. The more detailed experimental configuration around the sample and electromagnet are provided in Fig. S2 and Supplementary Note II.

The projected images of X-ray absorption (XAS) and magnetic (XMCD) contrast were collected by scanning the sample two-dimensionally in the $X$-$Z$ plane as a function of the rotation angle $\theta$ between -90$^\circ$ and +90$^\circ$ with a step of 5$^\circ$. The helicity-modulation technique with X-ray photon helicity switching at 37 Hz was used for lock-in detection of the dichroic signals with a high signal-to-noise ratio. The XAS and XMCD projections were recorded simultaneously. For reconstructing a three-dimensional (3D) XAS image, the standard algorithm of the algebraic reconstruction technique (ART)\cite{TomographyMath} was applied to 37 projection images collected at the angles from -90$^\circ$ to +90$^\circ$. To reconstruct a 3D magnetic image, a modified ART algorithm has been applied to XMCD projections. In the modified ART, the [001] component of local magnetization $m^c$ (in accord with the easy-axis uniaxial magnetic anisotropy of the present compound) was assumed and a correction for the $\cos \theta$ dependence of the XMCD amplitudes has been included\cite{TomographySuzuki}, then we obtained the 3D distribution of the $m^c ({\bf r})$ component of the magnetization vector of the sample. This reconstruction method was previously shown to reproduce the 3D structure of the internal magnetic domains in a GdFeCo disk, which has a strong uniaxial anisotropy, with a spatial resolution of 360 nm\cite{TomographySuzuki}.

\subsection*{Micromagnetic simulation.}

Micromagnetic simulation of 3D spatial distribution of vector magnetization ${\bf m}({\bf r}) \equiv {\bf m}_{\bf r}$ has been performed based on Landau-Lifshitz-Gilbert (LLG) equation with the magnetic Hamiltonian described as
\begin{equation}
\begin{split}
\mathcal{H}=&-J\sum_{\bf r} {\bf m}_{\bf r} \cdot ({\bf m}_{{\bf r}+{\bf \hat{a}}} + {\bf m}_{{\bf r}+{\bf \hat{b}}} + {\bf m}_{{\bf r}+{\bf \hat{c}}}) + D \sum_{\bf r} [{\bf \hat{a}} \cdot ({\bf m}_{\bf r} \times {\bf m}_{{\bf r}+{\bf \hat{a}}}) -  {\bf \hat{b}} \cdot ({\bf m}_{\bf r} \times {\bf m}_{{\bf r}+{\bf \hat{b}}})] \\
& - K_{\rm imp} \sum_{{\bf r} \in \Lambda} (m_{\bf r}^c)^2 - B\sum_{\bf r} m_{\bf r}^c. 
\end{split}
\end{equation}
${\bf \hat{a}}$, ${\bf \hat{b}}$ and ${\bf \hat{c}}$ are the unit vectors along the [100], [010] and [001] directions, and $m^c_{\bf r}$ is the [001] component of ${\bf m}_{\bf r}$. $J$, $D$, and $B$ represent the amplitude of exchange interaction, DM interaction with $D_{2d}$ symmetry, and external magnetic field along the [001] direction, respectively. The single-ion magnetic anisotropy $K_{\rm imp}$ is introduced on the random sites ${\bf r} \in \Lambda$ to represent the disorder in the crystals\cite{CurrentControl3D3}. For $\Lambda$, we use the random-number generator developed by M. Matsumoto and T. Nishimura (http://www.math.sci.hiroshima-u.ac.jp/~m-mat/MT/emt.html). Here, we employed the system size with $100 \times 50 \times 100$ sites and the parameter set $J=1, D=0.2, B=0.02$ and $K_{\rm imp}=0.2$ with the defect density $6\%$. In horizontal (vertical) direction, periodic (open) boundary condition is used. The length of local magnetic moment is fixed to $|{\bf m}_{\bf r}|=1$. (See Supplementary Note IV for the detail).

\section*{Data availability} 

The data presented in the current study are available from the corresponding authors on reasonable request.

\section*{Author contributions} S.S, M.S and T.O planned the project. S.S, M.S, M.I, T.O performed the magnetic X-ray tomography measurements. R.T and N.D.K prepared the samples. W.K performed the theoretical calculations. S.S wrote the manuscript with the support by M.S and T.O. All authors discussed the results and commented on the manuscript.

\section*{Acknowledge} The authors thank L. Peng, X. Z. Yu, N. Nagaosa, T. Arima and A. Kikkawa for enlightening discussions and experimental helps. This work was partly supported by Grants-In-Aid for Scientific Research (A) (grant nos 18H03685 and 20H00349) and Grant-in-Aid for Specially Promoted Research (15H05702) from JSPS, PRESTO (grant no. JPMJPR18L5) and CREST (grant no. JPMJCR1874) from JST, Asahi Glass Foundation and Murata Science Foundation. The synchrotron radiation experiments were performed with the approval of the Japan Synchrotron Radiation Research Institute (JASRI) (Proposal Nos. 2018A2067 and 2019B1173).

\section*{Additional information} Supplementary information is available in the online version of the paper. Reprints and permissions information is available online at www.nature.com/reprints. Correspondence and requests for materials should be addressed to S.S, M.S, or T.O. (email: seki@ap.t.u-tokyo.ac.jp, m-suzuki@spring8.or.jp, ono@scl.kyoto-u.ac.jp)

\section*{Competing financial interests}  The authors declare that they have no competing financial interests.

\newpage

\end{document}